\documentclass[prr,
aps,
amsmath,
amssymb,
twocolumn,
footinbib,
superscriptaddress
]{revtex4-2}
     \usepackage{url}
\usepackage{ragged2e}       
\usepackage{amssymb,amsmath,xcolor}
\usepackage[caption=false]{subfig}
\usepackage[makeroom]{cancel}
\usepackage[export]{adjustbox}
\usepackage{hyperref} 
\usepackage{subfig}
\usepackage{stackengine}
\usepackage{enumerate}
\usepackage{xr-hyper}
\usepackage{stackengine}
\begin{document}
\title{Coherent radiation of an electron bunch colliding with an intense laser pulse}
\author{E.~G. Gelfer}\email{egelfer@gmail.com}
\affiliation{ELI Beamlines facility, The Extreme Light Infrastructure ERIC,
Dolni Brezany 252 41, Czech Republic}
\author{A.~M.~Fedotov}\email{am\_fedotov@mail.ru}
\affiliation{National Research Nuclear University MEPhI, Moscow, 115409,  Russia}
\author{O.~Klimo}%\email{Ondrej.Klimo@eli-beams.eu}
\affiliation{ELI Beamlines facility, The Extreme Light Infrastructure ERIC,
Dolni Brezany 252 41, Czech Republic}
\affiliation{FNSPE, Czech Technical University in Prague, Prague, Czech Republic}
\author{S.~Weber}
\affiliation{ELI Beamlines facility, The Extreme Light Infrastructure ERIC,
Dolni Brezany 252 41, Czech Republic}

\begin{abstract}
We study the conditions for coherent radiation of an electron bunch driven by a counterpropagating strong pulsed electromagnetic plane wave. We derive the spectrum of the coherent radiation and show that it is emitted backwards with respect to the laser propagation direction and has a very narrow angular spread. We demonstrate that for a solid density plasma coherent radiation extends to frequencies up to hundreds of keV thereby enhancing the low-frequency part of the spectrum by many orders of magnitude. Our analytical findings are tested with 3D particle-in-cell simulations of an electron bunch passing through a laser pulse, clearly demonstrating how the coherence can essentially modify the observed radiation spectrum.
\end{abstract}

\maketitle

An electron quivering in a laser field acquires relativistic energy if the dimensionless field strength $a_0$ obeys
\begin{equation}
a_0=\frac{eE_0}{m\omega_Lc}\gtrsim1,
\end{equation}
where $E_0$ is the electric field amplitude, $\omega_L$ is the field frequency, $-e$ and $m$ are the electron charge and mass, $c$ is the speed of light. Radiation of the electron in such a regime is called the nonlinear inverse Thomson/Compton scattering (NTS/NCS) depending on whether or not the recoil of the electron due to emissions of individual photons can be neglected \cite{dipiazza_rev2012}. For brevity, we omit the word ``inverse'' in the following. 

Laser driven NTS and NCS serve as a source of bright X-- and gamma rays   \cite{phuoc_natphot2012,dromey_natphys2012,dromey_njp2013, nerush_pop2014, bashinov_epjst2014, rykovanov_jpb2014, geddes_nuclinstrb2015, yan_natphot2017, gu_commphys2018, zhu_apl2018, valialshchikov_prl2021, mironov_pra2021} with important applications in nuclear spectroscopy \cite{carman_nia1996,albert_prab2010,albert_prab2011,geddes_nib2015}, medicine \cite{carroll_ajr2003,weeks_medphys1997}, X-ray radiography \cite{quiter_jap2008,phuoc_natphot2012}, positron production via quantum electrodynamical (QED) cascades \cite{bell_prl2008,fedotov_prl2010,gelfer_pra2015, jirka_pre2016, samsonov_scirep2019, mironov_pra2021}, etc., see the review \cite{albert_ppcf2016}.  

The theory of NTS by a single electron in a monochromatic plane wave is well established~\cite{sarachik_prd1970, ritus1985,landau2,esarey_pre1993,salamin_pra1997}. However, when a laser irradiates a target constituted of many particles, the emitted radiation can be coherently enhanced. While the energy of incoherent radiation scales with the number of particles $N$ as $\mathcal{E}^{(N)}_{i}\sim N$, the coherent scaling would be $\mathcal{E}^{(N)}_{c}\sim N^2$. This possibility was explored and indeed confirmed for two nonrelativistic orbiting particles \cite{klepikov_ufn1985}, synchrotron radiation by electrons in circular accelerators \cite{schwinger1945,michel_prl1982,nakazato_prl1989,hirschmugl_pra1991,berryman_nimra1996}, self-amplified spontaneous emission by electrons in undulators \cite{madey_jap1971,kondratenko_partacc1980,emma_natphot2010,pellegrini_revmodphys2016}, production of light scalars due to the beyond-the-standard-model interactions \cite{dillon_prd2019} and the linear Thomson scattering \cite{schaap_njp2022}. Here we investigate, how the coherence manifests itself in the radiation of an electron bunch interacting with an intense laser pulse. 

In order to radiate coherently, the particles should move in an external field along nearby trajectories with close initial coordinates and momenta. The distance between the particles is prescribed by their density, in the current paper we focus solely on its impact on coherence assuming equal initial momenta. The effect of momentum spread (due to e.g. a nonzero temperature or uncertainty principle) will be neglected but was discussed in Ref.~\cite{angiogi_prl2018}. 

The effect of coherency depends on the radiated frequency. Radiation at frequencies above some threshold $\omega_c$ is incoherent and its spectrum is proportional to the single particle one, $d\mathcal{E}^{(N)}/d\omega=Nd\mathcal{E}^{(1)}/d\omega$, while for $\omega<\omega_c$ the spectrum is modified by coherency. For many-particle radiation spectra in accelerators \cite{michel_prl1982} and laser fields \cite{gonoskov_pre2015} it was suggested that the low frequency part of the spectra scales as $\omega^{-8/3}$ in contrast to $\omega^{1/3}$ as for a single particle \footnote{The considered frequencies are assumed still higher than the frequency of variation of the external field.}
%The coherency enhances radiation at lower frequencies stronger than at higher ones, hence modifies the radiation spectrum. For NTS this was pointed out in \cite{gonoskov_pre2015}. Many-particle radiation spectra in accelerators \cite{michel_prl1982} and laser fields \cite{gonoskov_pre2015} at low frequencies  $\omega$ often scale as $\omega^{-8/3}$ in contrast to $\omega^{1/3}$ as for a single particle \footnote{The considered frequencies are assumed still higher than the frequency of variation of the external field.}. 
It can be understood in the following way. If one assumes that only the particles separated by the distance shorter than the wavelength $\lambda$ of the emitted radiation radiate coherently, then the coherence enhances the radiation at certain frequency by the factor of the number of the particles inside the sphere of the radius $\lambda$. Consequently, the single particle synchrotron scaling $\omega^{1/3}$ \cite{ritus1985,esarey_pre1993} should be multiplied by $\omega^{-3}$, the factor proportional to the volume inside the sphere. 

The same ($\propto\omega^{-8/3}$) scaling  \cite{baeva_pre2006,dromey_natphys2012} was found for the relativistic oscillating mirror (ROM) mechanism of high harmonics generation by  reflecting laser pulses from solid targets \cite{bulanov_pop1994,gordienko_prl2004,baeva_pre2006,thaury_natphys2007} (note that ROM can also be created by a laser in a form of a plasma wave \cite{bulanov_prl2003,li_apl2014}). However, it was argued  \cite{brugge_pop2010,dromey_natphys2012,dromey_njp2013,boyd_prl2008,gonoskov_pre2011,edwards_scirep2020,bhadoria_pop2022} that the resulting spectra can differ if the competing mechanisms of high harmonics generation become important. 

In the present paper we consider the radiation of a relativistic electron bunch colliding head on with a short intense laser pulse. We determine the threshold frequency $\omega_c$ for coherence, derive the coherent part of the spectrum for $\omega<\omega_c$ and show that the reasoning leading to the scaling $\omega^{-8/3}$ for NTS does not work. We also demonstrate that the angular distributions of the coherent and incoherent radiation are different by that the coherent radiation is emitted opposite to the laser propagation direction and has a very narrow angular spread. In a wide range of laser and bunch parameters the energy of the coherent radiation surpasses the incoherent one by several orders of magnitude.

For simplicity, we consider circularly polarized laser pulse propagating along $x$ axis with vector potential $\mathbf{A}(\phi)=mca_0 g(\phi)/e\{0,-\sin\phi,\cos\phi\}$. Here $\phi=\omega_L(t-x/c)$ is the phase, $\omega_L$ is the frequency and $g(\phi)$ is the envelope of the pulse. According to \cite{jackson_book1999,landau2}, the angular resolved energy spectrum of the single particle radiation reads 
\begin{equation}\label{dE}
\frac{d\mathcal{E}^{(1)}}{d\omega d\Omega}=\frac{e^2\omega^2}{4\pi^2 c}\left|\int[\mathbf{n}\times[\mathbf{n}\times\mathbf{v}(t)]]e^{i\omega\left(t-\frac{\mathbf{n}\mathbf{r}(t)}{c}\right)}dt\right|^2,
\end{equation}
where $d\Omega=\sin\theta d\theta d\varphi$, $\omega$ and $\mathbf{n}=\{\cos\theta,\sin\theta\cos\varphi,\sin\theta\sin\varphi\}$ are the frequency and the direction of the emitted radiation, $\mathbf{v}(t)$ and $\mathbf{r}(t)$ are the velocity and coordinates of the particle. 

Let us consider the radiation of $N$ particles and assume that the initial momenta of all particles are equal. Then, changing the integration over time to the integration over phase $dt=\frac{mc\gamma}{\omega_L p_-} d\phi$ ($\gamma$ is the gamma factor, $p_-=\gamma mc-p_x$, $\mathbf{p}$ is the momentum, the laser propagates along $x$ axis), and taking into account that $\mathbf{p}_j(\phi)\equiv \mathbf{p}(\phi)$ are identical for all the particles and $\mathbf{r}_j(\phi)=\mathbf{r}_j^0+\mathbf{r}(\phi)$, where  $\mathbf{r}^0_j$, is the initial position of $j$'s particle,  we get
\begin{equation}\label{dEN}
\frac{d\mathcal{E}}{d\omega d\Omega}=\mathcal{C}\frac{d\mathcal{E}^{(1)}}{d\omega d\Omega},\quad \mathcal{C}=\left|\sum\limits_{j=1}^N e^{i\Phi_j}\right|^2,
\end{equation}
where $\Phi_j=\omega(x_j^0-\mathbf{n}\mathbf{r}_j^0)/c$. 

To evaluate the coherence factor $\mathcal{C}$ we decompose the sum into $N$ diagonal and $N(N-1)$ off-diagonal terms \cite{schiff_rsi1946}
\begin{equation}\label{C}
\mathcal{C}=N+\sum\limits_{j\neq k}^N e^{i(\Phi_j-\Phi_k)}=N(1-\alpha)+N^2\alpha,
\end{equation}
where
\begin{equation}\label{alpha}
\alpha=\left|\left<e^{i\Phi}\right>\right|^2=\frac{4 c^4 J_1^2\left(\frac{\omega R}{c}\sin\theta\right)\sin^2\left(\frac{\omega L}{c}\sin^2\frac{\theta}{2}\right)}{ R^2L^2\omega^4\sin^2\theta\sin^4\frac{\theta}{2}},
\end{equation}
$\left<\ldots\right>$ corresponds to the averaging over the particle position in the bunch, $J_1(\eta)$ is the Bessel function, $\theta$ is the angle between the wave vectors of the laser the  emitted radiation and we assume that the bunch is a cylinder of length $L$, radius $R$ and density $n$.

The two terms in the expression (\ref{C}) correspond to the incoherent ($\sim N$) and coherent ($\sim N^2$) contributions to the radiation. First, we note that the full coherence $\mathcal{C}=N^2$ is realized if either  $\omega R/c,\omega L/c\ll1$, or for the forward scattering $\theta=0$. In the first case  the whole bunch can be seen as a single particle with the charge $Ne$, while the radiated energy scales as charge squared, see Eq.~(\ref{dE}). In the case of forward scattering all the phases $\Phi_j=0$, and the radiation is indeed fully coherent. However, the emission of high frequencies in the forward direction is strongly suppressed by the single particle radiation term $d\mathcal{E}^{(1)}/d\omega d\Omega$ (see below).

Let us consider a large bunch, high frequencies ($\omega R/c,\omega L/c\gtrsim1$) and determine the conditions providing the coherency in the sense that the second (coherent) term in (\ref{C}) prevails over the first one. First of all, we note that in general the coherent contribution is strongly suppressed for high frequencies, since it is proportional to $\omega^{-5}$ (one additional power of $\omega$ appears because of the expansion of the Bessel function for a large argument). Nevertheless, coherence is enhanced in two particular directions, namely forward, which has been discussed already, and backward ($\theta=\pi$). Indeed, in the backward direction the radiation is coherent, if $N\alpha(\theta=\pi)\gg 1$, i.e.
\begin{equation}\label{omegac}
\omega\ll\omega_c=\frac{c}{L}\sqrt{N}=\omega_L\sqrt{\zeta\frac{n}{n_c}\frac{R^2}{\lambda_L L}},
\end{equation}
where $\zeta=\lambda_L/4 r_e$, $r_e=e^2/mc^2\approx2.8\cdot10^{-13}$ cm is the classical electron radius, and $n_c=m\omega_L^2/4\pi e^2$ is the plasma critical density. For the laser wavelength $\lambda_L=1\mu$m we have $\zeta\approx0.9\times 10^8$ and $n_c\approx 1.1\times 10^{21}$ cm$^{-3}$.

To obtain the spectrum $d\mathcal{E}/d\omega d\Omega$ one also needs $d\mathcal{E}^{(1)}/d\omega d\Omega$. The exact analytical expression for the single particle spectrum has been calculated for an infinite plane wave and a flat-top pulse \cite{ritus1985, landau2, sarachik_prd1970, esarey_pre1993, salamin_pra1997}, but is missing for realistic shapes of a laser pulse (e.g. Gaussian). Fortunately, $d\mathcal{E}^{(1)}/d\omega d\Omega$ can be evaluated in the forward and backward directions, where the coherence is most pronounced. 

Indeed, taking into account that the transverse to the laser propagation component of the velocity $\mathbf{v}_\bot=e\mathbf{A}/\gamma$, from Eq.~(\ref{dE}) we get
\begin{equation}\label{dE1}
\left.\frac{d\mathcal{E}^{(1)}}{d\omega d\Omega}\right|_{\theta=0,\pi}=\frac{e^4\omega^2}{4\pi^2c\omega_L^2p_-^2}\left|\int\mathbf{A}(\phi) e^{i\frac{\omega}{\omega_L}(\phi+\upsilon(\phi))}d\phi\right|^2,
\end{equation}
where $\upsilon(\phi)=0$ for $\theta=0$ and $\upsilon(\phi)=2\omega_Lx(\phi)/c$ for $\theta=\pi$. Therefore the spectrum of the forward scattering has the shape similar to the spectrum of the laser pulse, i.e. has a narrow peak around $\omega=\omega_L$ and is strongly suppressed for $\omega\gg\omega_L$ \cite{sarachik_prd1970}. 

\begin{figure}[t]
\includegraphics[width=0.47\textwidth,valign=c]{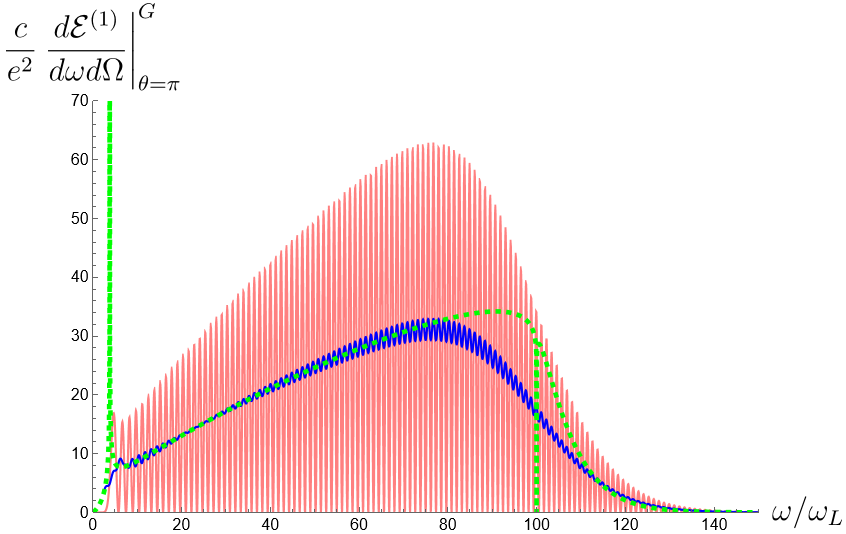}
\caption{Backscattered radiation of a single particle colliding with a laser pulse with Gaussian temporal envelope. Pink and blue -- numerical calculation and its average over the fast oscillations, dashed green curve -- analytical estimate. Vertical green lines at $\omega\approx4\omega_L$ and  $\omega=100\omega_L$ correspond to the margins of the interval (\ref{interv}). $a_0=5, p_-=10 mc, \omega_L T=5\pi$. }\label{fig_backsc}
\end{figure}

For the backward scattering, $\theta=\pi$, one can evaluate the integral employing the stationary phase approximation \cite{kharin_pra2016,seipt_lasphys2013}. 
In particular, for the Gaussian envelope $g(\phi)=e^{-\phi^2/(\omega_L T)^2}$ ($T$ is the pulse duration) of the laser one obtains
\begin{equation}\label{dE1G}
\left.\frac{d\mathcal{E}^{(1)}}{d\omega d\Omega}\right|_{\theta=\pi}\approx \frac{e^2}{4\pi c}\frac{\omega T}{\nu(\omega)}
\begin{cases}
1, &\omega<\omega_*\\
e^{-\frac{(\omega-\omega_*)\pi\omega_L T}{\nu(\omega)\omega_*}}, &\omega>\omega_*
\end{cases},
\end{equation}
%where $\nu(\omega)=2\phi_1/(\omega_L T)=\sqrt{2\ln\frac{\omega a_0^2}{|\omega-\omega_*|}}$. Hence for the Gaussian laser pulse the spectrum of the backwards scattered radiation starts from $\omega>\omega_*/(1+a_0^2)$, then it scales almost linearly with the frequency until $\omega<\omega_*$ and exponentially decays for $\omega\gg\omega_*$. 
where $\nu(\omega)=\sqrt{2\ln\frac{\omega a_0^2}{|\omega-\omega_*|}}$. %Hence for the Gaussian laser pulse 
The spectrum of the backward scattered radiation  scales almost linearly with the frequency in the interval
\begin{equation}\label{interv}
\omega_L\xi^2<\omega<\omega_*\equiv\omega_L\left(\frac{p_-}{mc}\right)^2,\quad \xi=\frac{p_-}{mc\sqrt{1+a_0^2}},
\end{equation}
and is strongly suppressed outside this interval. Note that for simplicity Eq.~(\ref{dE1G}) is averaged over fast oscillations. It perfectly fits the average backward spectrum obtained by numerical evaluation of  (\ref{dE1}), everywhere except for the vicinities of the interval (\ref{interv}) endpoints, see Fig.~\ref{fig_backsc}.

Setting $\theta=\pi$ in the second term of (\ref{C}) and combining it with (\ref{dE1G}) we get the estimate for the coherent backward scattering
\begin{equation}\label{dEdomdoc}
%\left.\frac{d\left(\frac{\mathcal{E}}{mc^2}\right)}{d\left(\frac{\omega}{\omega_L}\right)d\Omega}\right|_{\theta=\pi}\approx\frac{\pi^3\zeta}{2}\left(\frac{n}{n_c}\right)^2\left(\frac{ R}{\lambda_L}\right)^4\frac{\omega_L T}{\nu(\omega)}\frac{\omega_L}{\omega}\sin^2\frac{\omega L}{c},
\left.\frac{d\mathcal{E}}{d\omega d\Omega}\right|_{\theta=\pi}\approx\frac{\pi^2\zeta}{4}\left(\frac{n}{n_c}\right)^2\left(\frac{ R}{\lambda_L}\right)^4\frac{\omega_L T}{\nu(\omega)}\frac{mc^2}{\omega},%\sin^2\frac{\omega L}{c},
\end{equation}
for $\omega$ inside the interval (\ref{interv}) and $\omega<\omega_c$. Here we averaged one more oscillating factor $\left<\sin^2\frac{\omega L}{c}\right>=1/2$. As usual, the coherent radiation energy scales as $\propto n^2$ rather than $\propto n$.

\begin{figure}[t]
\topinset{(a)}{\subfloat{\includegraphics[width=0.49\linewidth,valign=c]{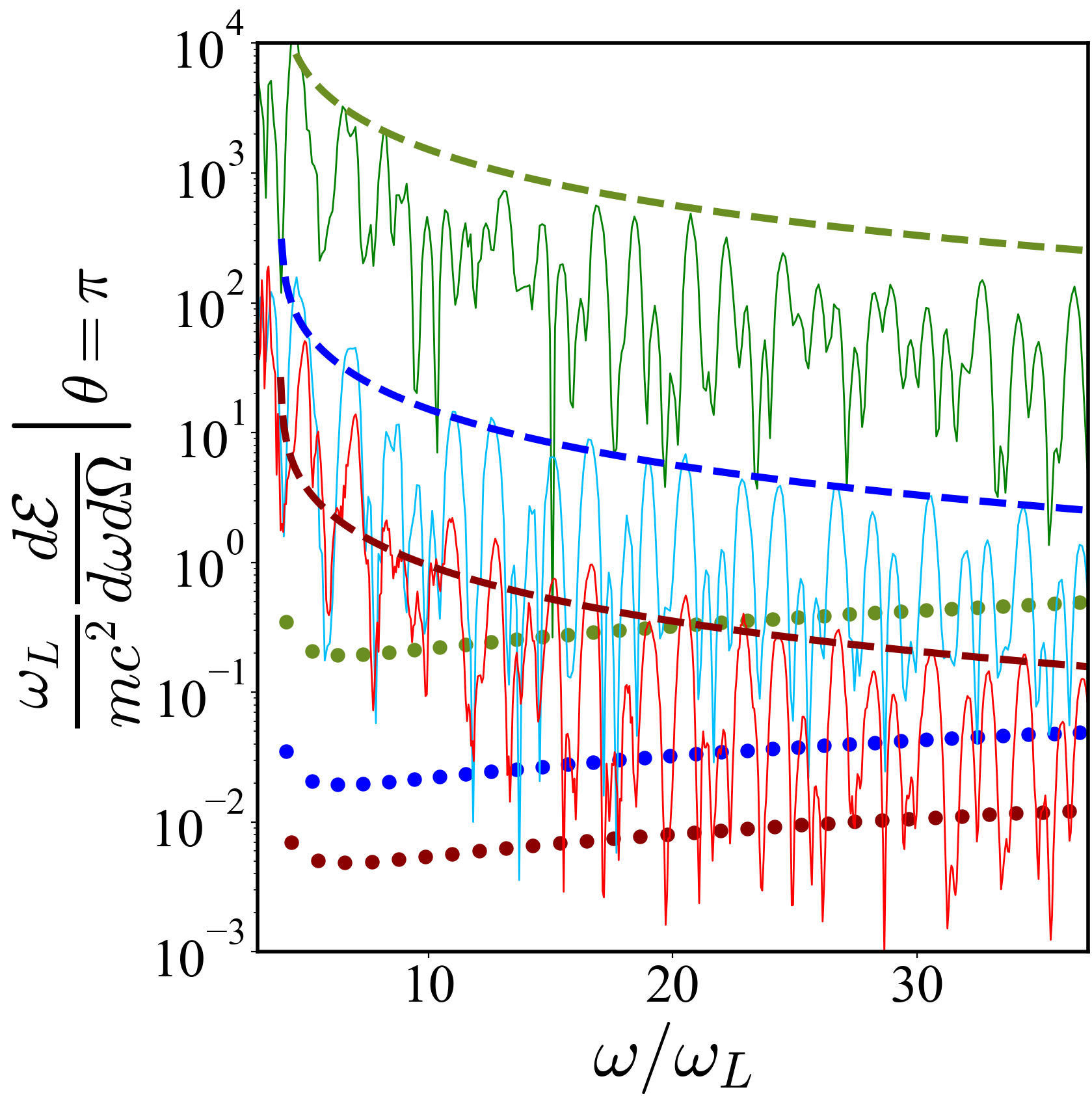}}}{0.25cm}{-1.8cm}
%\stackinset{r}{0.1cm}{t}{0.5cm}{\includegraphics[width=0.13\linewidth,valign=c]{dEdodO_x28_omega15.png}\includegraphics[width=0.13\linewidth,valign=c]{dEdodO_x28_omega36.png}}
\topinset{(b)}{\subfloat{\includegraphics[width=0.49\linewidth,valign=c]{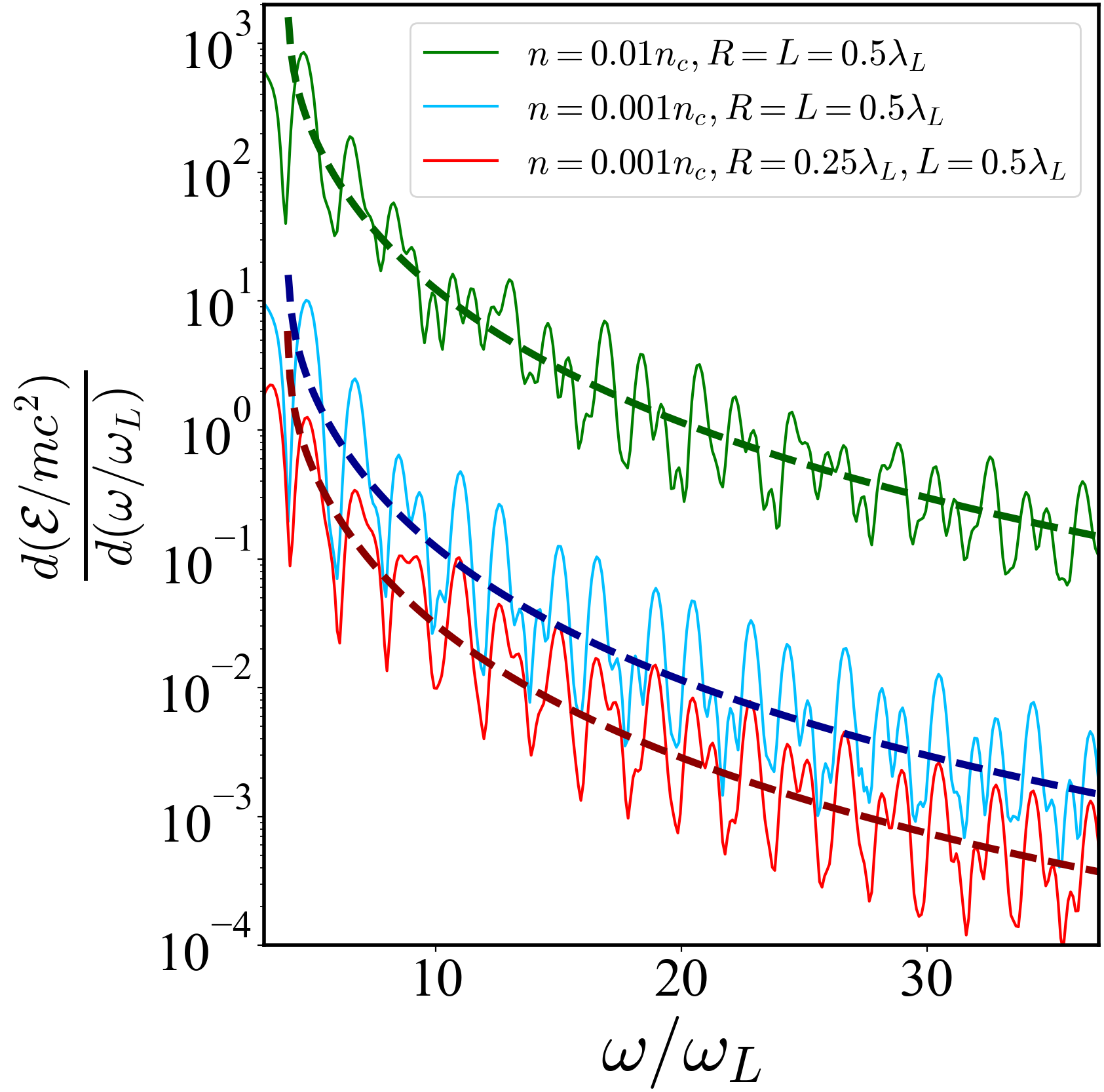}}}{0.5cm}{-1.65cm}\\
\topinset{(c)}{\subfloat{\includegraphics[width=0.49\linewidth,valign=c]{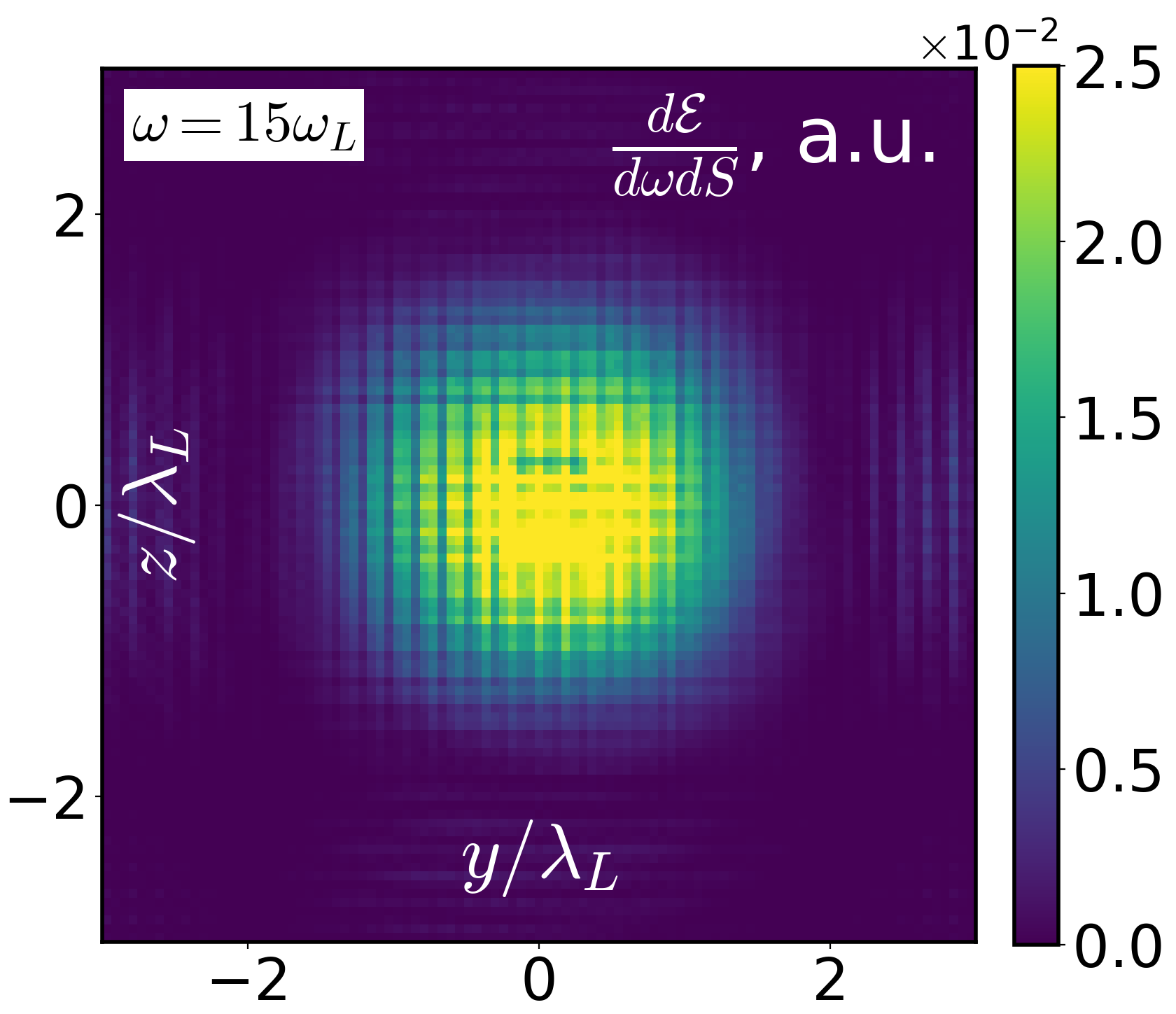}}}{0.2cm}{-1.3cm}
\topinset{(d)}{\subfloat{\includegraphics[width=0.49\linewidth,valign=c]{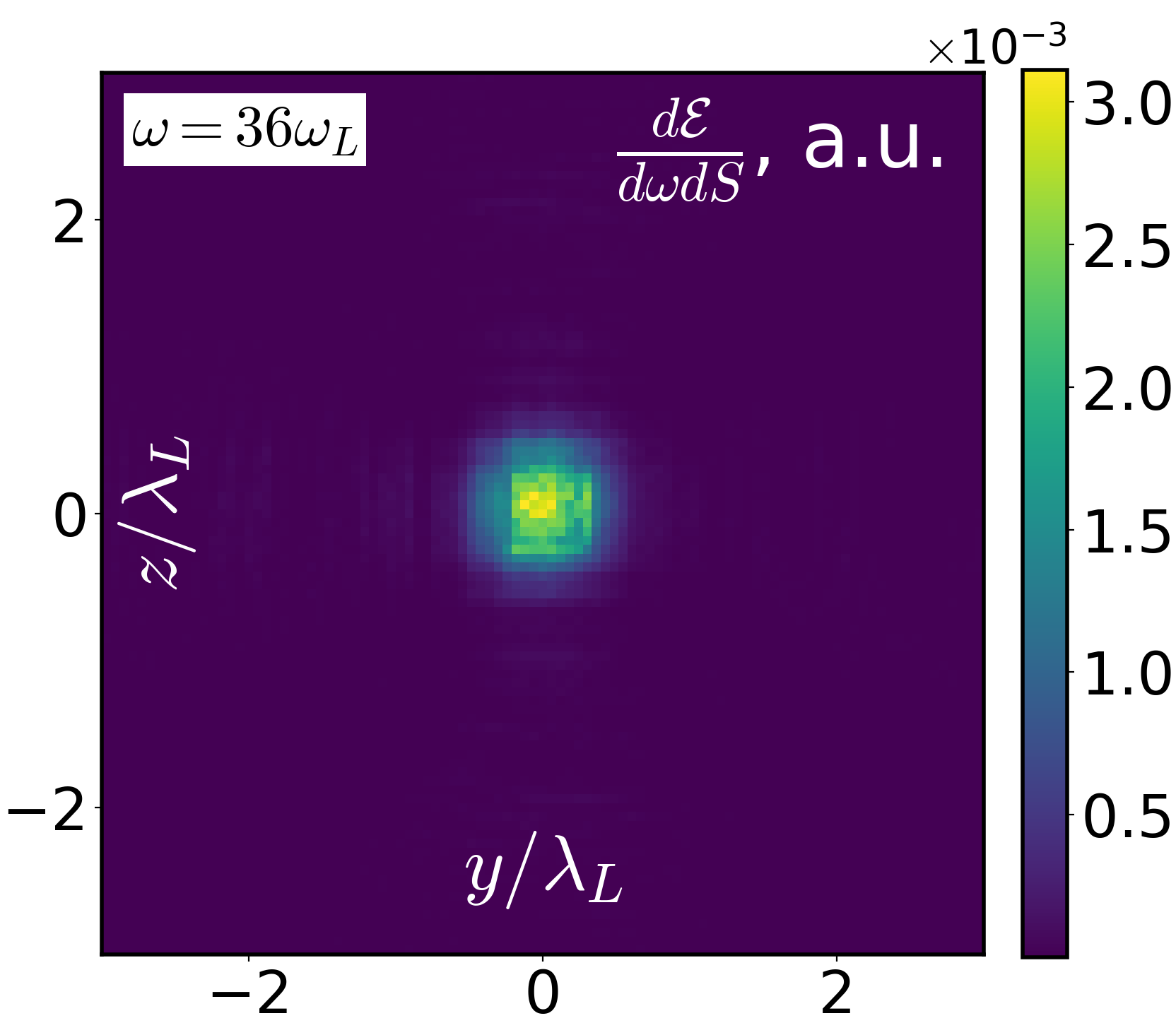}}}{0.2cm}{-1.3cm}
\caption{Backward (a) and angular integrated (b) radiation spectra of an electron bunch colliding with a laser pulse with Gaussian temporal envelope for various bunch densities and sizes; $a_0=\gamma=5 \ [p_-\approx10mc], \omega_LT=5\pi, \lambda_L=1\mu$m, other parameters are given at the plot (b) legend. Solid lines -- results of 3D numerical simulations using PIC code SMILEI, dashed lines -- analytical estimates Eq.~\eqref{dEdomdoc} and \eqref{dEdoc}, dotted lines -- incoherent analytical estimates ($\mathcal{C}=N$). (c) and (d) -- radiation distribution $d\mathcal{E}/d\omega dS$ [$dS=dydz$]  in the $yz$ plane at $x=28\lambda_L$ for the parameters of the blue curve and $\omega=15\omega_L$ (c), $\omega=36\omega_L$ (d). }\label{fig_spectsim}
\end{figure}

Let us briefly discuss the main features of the coherent radiation spectrum. First, it is proportional to the squared transverse cross section of the bunch, but (after averaging) is independent of the longitudinal size. % only through oscillatory factor $\sin^2\frac{\omega L}{c}$. 
This is natural, because the phases $\Phi_j$ in the backward direction are independent on particles positions in the transverse planes, but $e^{i\Phi_j}$ oscillate with longitudinal position of the particle. 

Second, the radiation is directed backward, it is narrow with the angular spread inversely proportional to the frequency $\Delta\theta\sim c/\omega R$.  The latter can be derived from the condition that the argument of the Bessel function in (\ref{C}) $\omega R\sin\theta/c\lesssim 1$. In contrast, the incoherent radiation is emitted along the cone with the opening angle $\theta_0=2\arctan\xi$, and its angular spread $\Delta\theta\sim\frac{\xi(\omega/\omega_L)^{-1/3}}{(1+\xi^2)^{2/3}}$ \cite{esarey_pre1993}.

Third, consider the total (integrated over the solid angle) spectrum
\begin{equation}\label{dEdoc}
\frac{d\mathcal{E}}{d\omega}\approx \frac{\pi\zeta}{4}\left(\frac{n}{n_c}\frac{R}{\lambda_L}\right)^2\left(\frac{\omega}{\omega_L}\right)^{-3}\frac{mc^2 T}{\nu(\omega)},
\end{equation}
%where we averaged over the oscillations of $\left<\sin^2\frac{\omega L}{c}\right>=1/2$. 
which is obtained under the assumption that the angular spread $\Delta\theta\sim c/\omega R$ determined by the coherence factor (\ref{C}) is narrow and that $d\mathcal{E}^{(1)}/d\omega d/\Omega$ remains constant for $\pi-\Delta\theta<\theta\leq\pi$. 
Spectrum (\ref{dEdoc}) scales with the frequency as $\omega^{-3}/\nu(\omega)$, which differs from single particle spectrum in an infinite wave  ($\omega^{1/3}$) \cite{ritus1985,esarey_pre1993}, as well as from the previous estimates of coherent NTS ($\omega^{-8/3}$) \cite{gonoskov_pre2015} and other mechanisms of the generation of coherent radiation, e.g. ROM,  \cite{bulanov_pop1994,gordienko_prl2004,baeva_pre2006,thaury_natphys2007,bulanov_prl2003,li_apl2014}.

%Finally, 

To check the analytical predictions we performed three-dimensional (3D) simulations of an electron bunch colliding head on with a laser pulse with the particle-in-cell (PIC) code SMILIEI \cite{derouillat_cpc2018}. In modern PIC codes the quantum and incoherent part of radiation by charged particles is implemented as emission of individual photons via Monte-Carlo event generators  \cite{gonoskov_pre2015}. Between the emission events particles are moving along the classical trajectories thus forming currents, which produce classical radiation. %Though modern PIC-codes are capable of describing radiation emitted in a quantum regime \cite{gonoskov_pre2015}, its implementation so far totally ignores all the collective effects, including possible coherent amplification. Being also interested in a low-frequency range where the classical description works well, we therefore analyzed the output of the built-in Maxwell solver.

In our  simulations quantum emission is turned off, electrons initially are moving to the right, momenta of all particles are equal to $p_x=5mc$. The bunch has the cuboid shape; the initial density of the bunch is uniform. Laser is a pulsed plane wave  with $a_0=5$ propagating to the left. The size of the simulation box is $30\lambda_L\times6\lambda_L\times 6\lambda_L$, the particles cross the peak of the laser field at $x\approx12.5\lambda_L$. The spacial resolution is 256 cells per wavelength in all directions, the temporal resolution is 512 time steps per laser period. Each cell with plasma contains 5 macroparticles.  

The results of simulations are presented in Fig.~\ref{fig_spectsim}. %They correspond to $\xi\approx 2>1$, for which radiation is mostly emitted along the $x$ axis. 
From simulation data we extract the field values at the distant plane $x=28\lambda_L$, perform their fast Fourier transform and combine them into the Fourier transformed Poynting vector, which is proportional to $d\mathcal{E}/d\omega d\Omega$. The approximate match of the solid (simulation) and dashed (estimated analytically) curves for different densities and bunch sizes is a good indication of the relevance of the scalings (\ref{dEdomdoc}), (\ref{dEdoc}) in density, frequency and size of the bunch. Spatial distributions of the radiation in a transverse plane also confirm the  narrowing with the increase of the frequency. 
%in density ($\propto n^2$), frequency ($\propto\omega^{-1}/\nu(\omega)$) and size of the bunch ($\propto R^4$). Spatial distributions of the radiation in a transverse plane also confirm the  narrowing with the increase of the frequency. 

PIC approach, however, has severe limitations, which currently make it challenging to simulate the effect of coherent amplification directly for more extreme parameters. First, the Maxwell solver resolves only frequencies obeying $\omega\ll d\omega_L$, where $d$ is the number of grid cells per wavelength. Second, in a PIC code the particles are combined into macroparticles densely populating grid cells and radiating coherently within each cell (this, in particular, means that so computed radiation can scale as $\propto\omega^{-1}/\nu(\omega)$ even for $\omega>\omega_{c}$). %An alternative approach \cite{quin_arxiv2023} relies on the numerical solving of the equations of real particles motion and following numerical evaluation of the radiation spectrum provided by the total current of the bunch. This is, however, numerically expensive for treating large (with respect to the emitted wavelengths) bunches consisting of many particles and providing stronger effect of coherence. 

\begin{figure}[t]
\topinset{(a)}{\subfloat{\includegraphics[width=0.49\linewidth,valign=c]{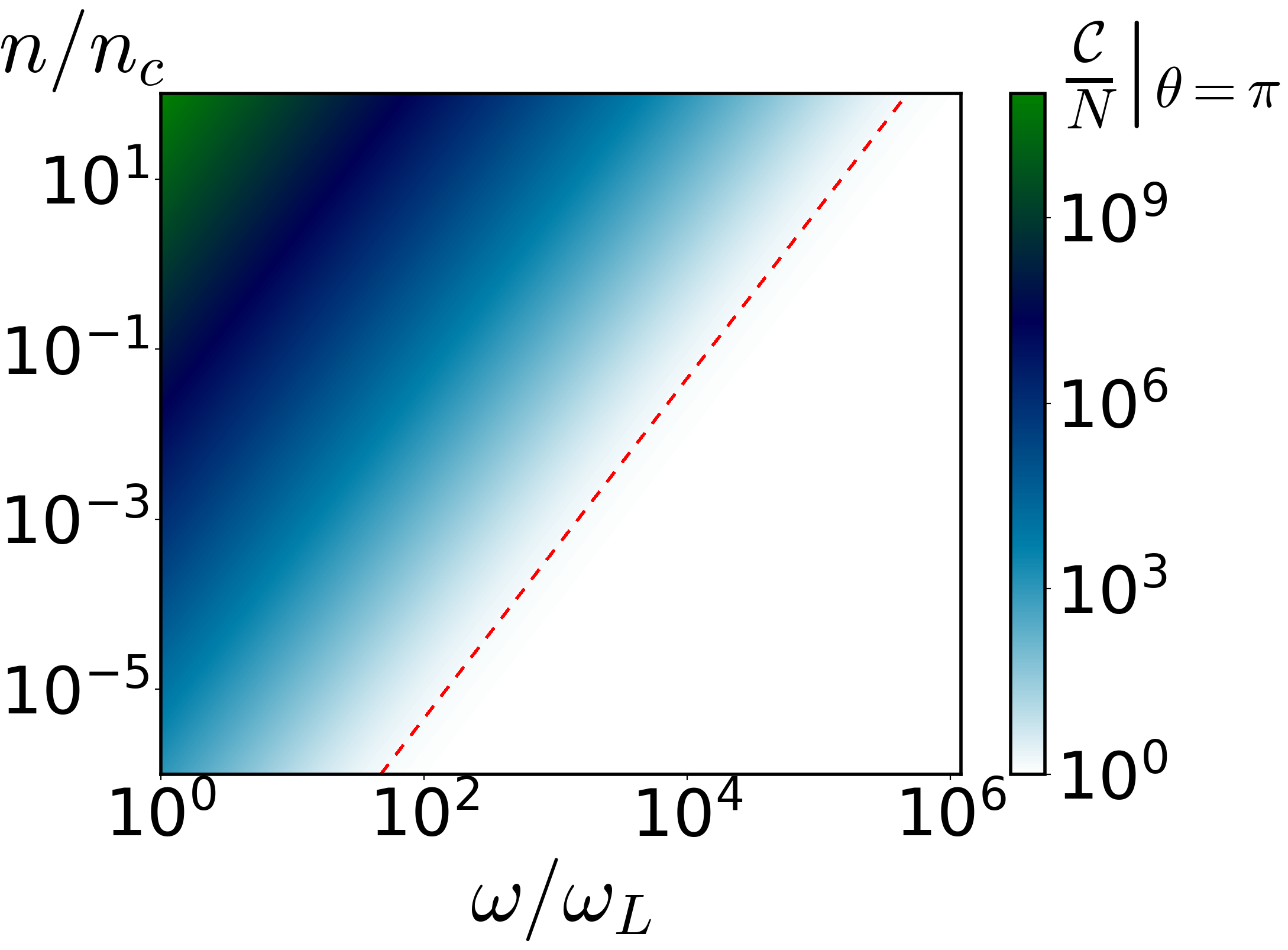}}}{0.7cm}{0.73cm}
\topinset{(b)}{\subfloat{\includegraphics[width=0.49\linewidth,valign=c]{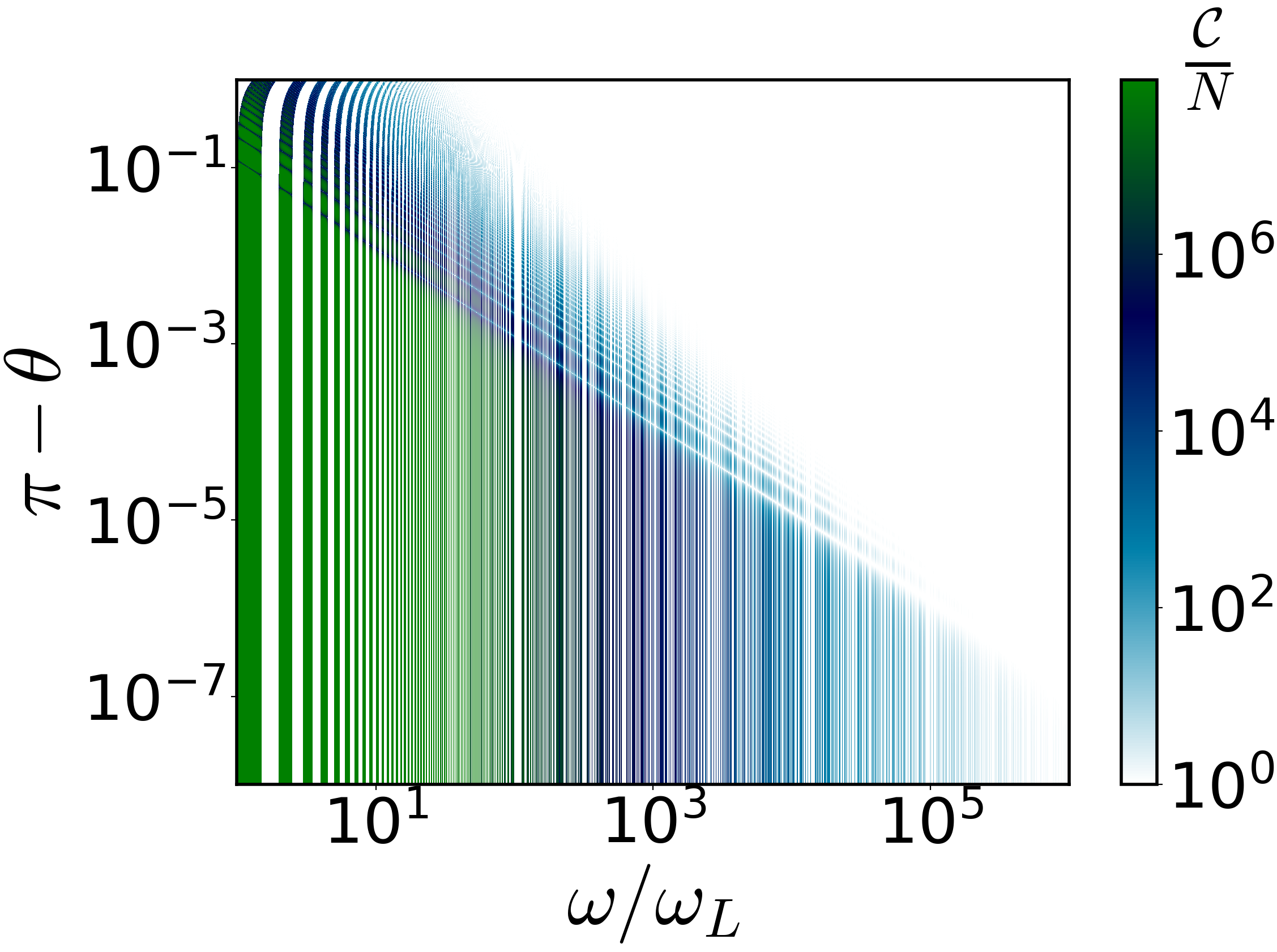}}}{0.95cm}{1.2cm}
\caption{Radiation enhancement factor $\mathcal{C}/N$ for backscattering (a) and $n=100 n_c$ (b); red dashed line corresponds to the estimate (\ref{omegac}) and separates the regions of coherent [$\mathcal{C}>N$] and incoherent [$\mathcal{C}=N$] radiation; $R=5\lambda_L$, $L=\lambda_L=1\mu$m. }\label{fig_cohest}
\end{figure}

Therefore, in analysing the parameters for which the coherent amplification can be revealed in laser--plasma interactions, we  rely mostly on extrapolation of our estimates. Consider $\mathcal{C}/N$, which corresponds to the coherent enhancement of the radiated energy compared to the incoherent radiation of $N$  particles. Its dependence on the radiated frequency $\omega$, density $n$ (with the upper limit $10^2$ corresponding to the solid state density) and the angle of radiation is illustrated in Fig.~\ref{fig_cohest}. Panel (a) corresponds to the backward scattering, and in this case the coherency affects the frequencies up to hundreds of keV range. In the low frequency limit the enhancement factor $\mathcal{C}/N$  exceeds $10^{10}$. Panel (b) corresponds to a solid density bunch ($n=100 n_c$) and demonstrates that at high frequencies coherency leads to the narrowing of the angular spread (up to $\mu$rad for $\omega\sim100$ keV). Note that at Fig.~\ref{fig_cohest}~(a) we averaged over the oscillations of $\sin^2(\omega L/c)$, that is  why it does not have strips structure similar to Fig.~\ref{fig_cohest}~(b).

According to \cite{sarachik_prd1970,esarey_pre1993} $\omega_s\sim a_0^3(1+\xi^2)$  is the frequency corresponding to the maximum of the single particle NTS spectrum, and since $\omega_s>\omega_*\geq\omega_c$ the radiation at the maximal frequencies is always incoherent, if $a_0\gg 1$. However, the  total energy $\mathcal{E}_{c}$ of the coherent part of the spectrum can be still higher than of incoherent one $\mathcal{E}_{i}$ because of the strong amplification of the lower frequencies, see Fig.~\ref{fig_cohest}.

\begin{figure}[t]
\topinset{(a)}{\subfloat{\includegraphics[width=0.49\linewidth,valign=c]{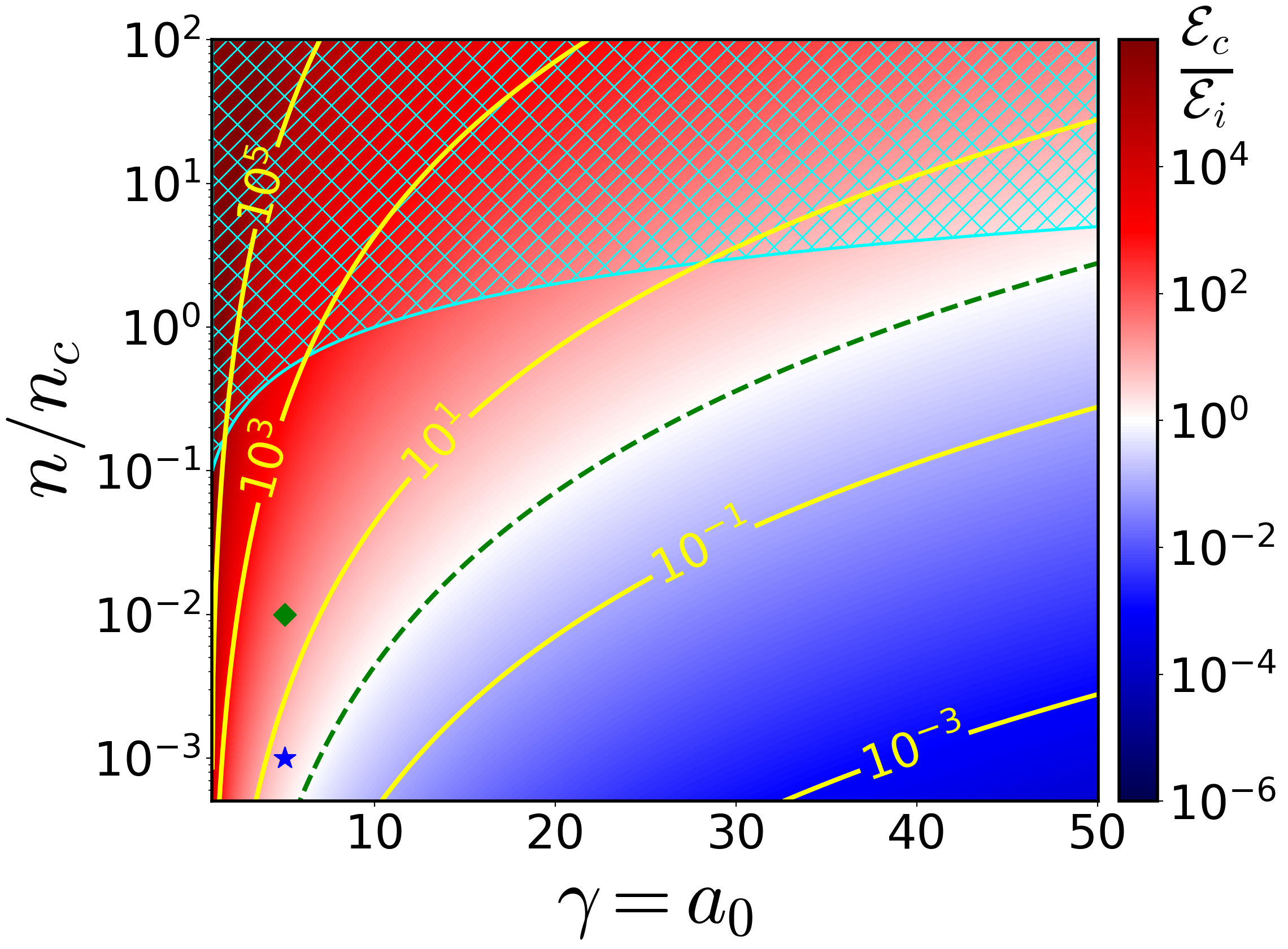}}}{0.2cm}{-2cm}
\topinset{(b)}{\subfloat{\includegraphics[width=0.49\linewidth,valign=c]{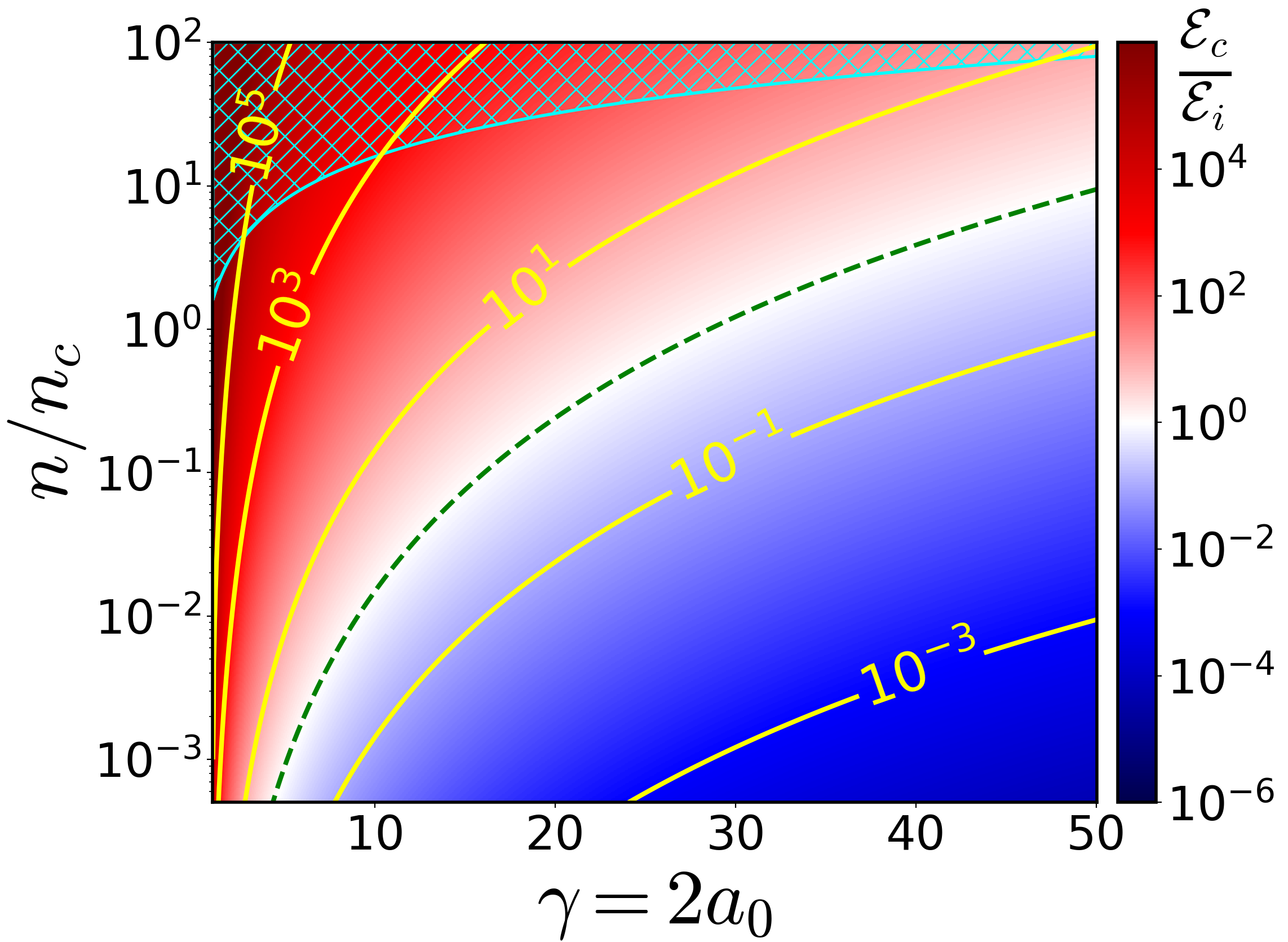}}}{0.5cm}{-2cm}
\caption{Ratio of the energies of coherent and incoherent parts of the spectrum vs the bunch density and gamma factor/laser strength for the cases $\gamma=a_0$ [$\xi\approx 2$] (a) and $\gamma=2a_0$ [$\xi\approx4$] (b). Green dashed lines represent $\mathcal{E}_c=\mathcal{E}_i$, hatched areas correspond to $\mathcal{E}_c<\mathcal{E}_b$, where our estimates are not quantitatively accurate, see Eq.~(\ref{Ecb}); blue star and green diamond correspond to the blue and green curves at Fig.~\ref{fig_spectsim}. $L=\lambda_L/2$, $\lambda_L=1\mu$m.  }\label{fig_eemap}
\end{figure}

Indeed, to estimate $\mathcal{E}_i$ we note that for $a_0\gg 1$ the major contribution to the total electromagnetic energy radiated by a single particle comes from the high frequencies $\omega\sim\omega_s$ and therefore $\mathcal{E}_{i}\sim N\mathcal{E}^{(1)}$, where $\mathcal{E}^{(1)}\sim e^2a_0^4\omega_L^2 T(1+\xi^2)/c$ is the estimate \cite{esarey_pre1993} for the total energy radiated by a single particle\footnote{Note that the expression for the total energy in \cite{esarey_pre1993} written in the paragraph below Eq.~(71b) should be corrected by the factor $\sqrt{M_0}/(1+M_0)$ ($M_0=\xi^2$ in our notations). Indeed, for integration Eqs. (69), (71a) over the solid angle one has to account for $\sin\theta$ in $d\Omega=\sin\theta d\theta d\varphi$ and $\sin\theta_0=2\sqrt{M_0}/(1+M_0)$, see Eq.~(66). }. On the other hand, integration of Eq.~(\ref{dEdoc}) over the interval (\ref{interv}) provides $\mathcal{E}_{c}\sim \zeta\omega_L T\left(\frac{n}{n_c}\frac{R}{\xi^2\lambda_L}\right)^2mc^2$ with the main contribution coming from the lower limit (note that our approach assumes that the radiated wavelength is shorter than the bunch, implying that $\xi>\sqrt{\lambda_L/R}$ and hence $a_0\lesssim(p_-/mc)\sqrt{R/\lambda_L}$). Therefore
\begin{equation}\label{Eci}
\frac{\mathcal{E}_{c}}{\mathcal{E}_{i}}\sim\frac{\zeta}{1+\xi^2}\frac{n}{n_c}\frac{\lambda_L}{L}\left(\frac{p_-}{mc}\right)^{-4},
\end{equation}
meaning that for $a_0=\gamma=5$ (parameters corresponding to Fig.~\ref{fig_spectsim}) $\mathcal{E}_c>\mathcal{E}_i$ already for $n<10^{-3}n_c$, see Fig.~\ref{fig_eemap}~(a), where blue star and green diamond correspond to the blue and green curves from Fig.~\ref{fig_spectsim}, and for these cases PIC simulations confirm the estimate (\ref{Eci}). Moreover, at the critical density the major part of the radiation is coherent for $\gamma\sim a_0\sim 40$, which corresponds to the intensity $I\sim4\times10^{21}$ W/cm$^2$, while  for $\gamma=2a_0=5$ we estimate $\mathcal{E}_c/\mathcal{E}_i\sim 10^3$, see Fig.~\ref{fig_eemap}~(b). This means that (i) the coherent part of the spectrum can be much more energetic than the incoherent one and (ii) neglecting the coherency can result in a strong underestimation of the total amount of radiation emitted by the bunch. 

In addition, an energy loss due to radiation can modify particles trajectories (radiation friction  \cite{landau2,dipiazza_rev2012}). This in turn can affect the radiation pattern. In our estimates we ignored this effect, assuming that the bunch energy $\mathcal{E}_b=N\gamma mc^2\gg \mathcal{E}_c$, or 
\begin{equation}\label{Ecb}
\gamma\gg(ncT/n_cL)^{1/5}a_0^{4/5}.
\end{equation}
 
To conclude, we demonstrated that the coherence can significantly modify the radiation spectrum of an electron bunch interacting with an intense laser pulse. The low frequency part of the spectrum is enhanced, the maximal frequency of the coherently enhanced radiation depends on the bunch  density and size, the angular spread of the coherent radiation is inversely proportional to the frequency and the total radiated energy can be strongly enhanced. These findings are essential for correct description of experiments with laser irradiated dense electron bunches. In particular, we identified the interaction regime, which is accessible at existing laser facilities and provides orders of magnitude dominance of the coherent radiation over the incoherent one. 
%Eq.~(\ref{Eci}) determines the condition providing the dominance of the coherent radiation over the incoherent one. 
Moreover, according to our preliminary simulations, the amplification of radiated energy can  also enhance the radiation friction force acting on  electrons in a dense bunch as compared to that in vacuum or in rare plasma \cite{landau2,bulanov_ppr2004,dipiazza_rev2012,fedotov_pra2014,gonoskov_prl2014,ji_prl2014,cole_prx2018,poder_prx2018,gelfer_scirep2018,gelfer_ppcf2018,gelfer_njp2021}. This implies that radiation friction in dense electron bunches might become essential at considerably lower laser intensities than commonly accepted.

The authors are grateful to Myron Sukhanov for the help with numerical simulations and to Mickael Grech, Daniel Seipt, Nina Elkina, Antonino Di Piazza, Michael Quin and Maxim Malahov for valuable discussions. A.M.F. was supported by the MEPhI Program Priority 2030 and by the Russian Science Foundation (Grant No. 20-12-00077, calculation of the coherence enhancement factor). The numerical simulations were performed using the code SMILEI and the resources of the ELI ERIC SUNRISE cluster.

%

%\bibliography{lit_doi2}

\end{document}